\def\year{2023}\relax
\documentclass[letterpaper]{article} 
\usepackage{aaai22}  
\usepackage{times}  
\usepackage{helvet}  
\usepackage{amsmath}
\usepackage{amssymb}
\usepackage{amsfonts}
\usepackage{courier}  
\usepackage[hyphens]{url}  
\usepackage{graphicx} 
\usepackage{natbib}  
\usepackage{caption} 
\DeclareCaptionStyle{ruled}{labelfont=normalfont,labelsep=colon,strut=off} 
\usepackage{algorithm}
\usepackage{algorithmic}

\usepackage{newfloat}
\usepackage{listings}
\floatstyle{ruled}
\newfloat{listing}{tb}{lst}{}
\floatname{listing}{Listing}
\title{Learning bias corrections for climate models using deep neural operators}
\author{
     Aniruddha Bora\equalcontrib\textsuperscript{\rm 1},
    Khemraj Shukla\equalcontrib\textsuperscript{\rm 1}, Shixuan Zhang\textsuperscript{\rm 2}, Bryce Harrop\textsuperscript{\rm 2}, Ruby Leung\textsuperscript{\rm 2} and George Em Karniadakis\textsuperscript{\rm 1} \\
}
\affiliations{
    \textsuperscript{\rm 1} Division of Applied Mathematics, Brown University,\\
   
    170, Hope Street Providence, RI-02906\\
    george\_karniadakis@brown.edu
\textsuperscript{\rm 2}Pacific Northwest National Laboratory\\

    902 Battelle Blvd, Richland, WA 99354\\
    
}

\usepackage{bibentry}

\begin{document}

\maketitle

\begin{abstract}
Numerical simulation  for climate modeling resolving all important scales is a computationally taxing process. Therefore, to circumvent this issue a low resolution simulation is performed, which is  subsequently corrected for bias using reanalyzed data (ERA5), known as nudging correction. The existing implementation for nudging correction uses a relaxation based method for the algebraic difference between low resolution and ERA5 data. In this study, we replace the bias correction process with a surrogate model based on the Deep Operator Network (DeepONet). DeepONet (Deep Operator Neural Network) learns the mapping from the state before nudging (a functional) to the nudging tendency (another functional). The nudging tendency is a very high dimensional data albeit having many low energy modes. Therefore, the DeepoNet is combined with a convolution based auto-encoder-decoder (AED) architecture in order to learn the nudging tendency in a lower dimensional latent space efficiently. The accuracy of the DeepONet model is tested against the nudging tendency obtained from the E3SMv2 (Energy Exascale Earth System Model) and shows good agreement. The overarching goal of this work is to deploy the DeepONet model in an online setting and replace the nudging module in the E3SM loop for better efficiency and accuracy.
\end{abstract}

\section{Introduction}
The atmospheric circulation is characterized by complex multiscale physical processes, hence posing a lot of challenges for numerical simulations.  These simulated processes may exhibit inconsistencies and significant deviations from the true states and are known as  ``biases". They are developed due to limited spatial resolution in climate models and inaccurate physical parameterizations in comparison to the observed atmospheric states. The limited resolution simulations are less taxing on computational resources, but their biases limit their usefulness. This trade off results into larger  uncertainties in  climate predictions. In general, such biases are removed using post processing steps, such as model output statistics for weather forecasting \cite{glahn1972use}, ensemble bias correction for seasonal prediction \cite{arribas2011glosea4}, \cite{stockdale1998global}, etc. In existing implementations of the online bias correction approach, the bias correction is based on a data driven approach with lower generalization  capabilities \cite{watt2021correcting}. In this work, we propose a neural operator based surrogate to learn the bias correction with a larger generalization capability. In particular, we utilize  Deep Operator Network (DeepONet) \cite{lu2021} by augmenting the existing neural network architecture with autoencoders \cite{oommen2022learning}.   

\noindent 
\section{Methodology}
In this section, we will discuss the data generation, methodology to train DeepONet, and the architecture details of the proposed surrogate model.

\subsection{Data generation}
The time evolution of a physical quantity $X_{m}$ in the Energy Exascale Earth System Model-E3SM \cite{doecode_10475} is given by \cite{sun2019impact} :
\begin{equation}\label{eq1}
    \frac{\partial X_{m}}{\partial t} = F(X_{m}), 
\end{equation}where $X_{m}$ is the state variable. The nudging tendency for the state variable is given by
\begin{equation}\label{eq2}
\left(\frac{\partial X_{m}}{\partial t}\right)_{\mathrm{n} d g}= \begin{cases}\frac{X_{p}\left(t_{p}\right)-X_{m}(t)}{\tau}, & t = t_{ i}, \\ 0, & t \neq t_{i},\end{cases}
\end{equation} where $t_{i}$ is the time at $i^{th}$ hour when the $ERA5$ data is available. The data used for training and testing the DeepONet are generated by performing EAMv2 simulations, which are nudged towards the reference data sets (i.e. 3-hourly ERA5 reanalysis data \cite{hersbach2020era5}). This time-dependent nudging tendency is determined by the model $X_{m}$ and reference state $X_{p}$ as well as the nudging relaxation time scale $\tau$. This is shown in \eqref{eq2}. Nudging tendencies are calculated at every 30 minutes following \eqref{eq2}. In \eqref{eq2}, $X_{p}$ is obtained by linearly interpolating the 3-hourly ERA5 reanalysis data to the model times. In this study, the EAMv2 simulations are nudged only for U (zonal wind) and V (meridional wind) and we have used three data sets, which are (i) data before nudging, (ii) nudging tendency and (iii) ERA5 reanalyzed data-set. These data sets, in relation to each other, are  expressed as, 
\begin{equation}\label{eq3}
\left(\frac{\partial X_{m}}{\partial t}\right)_{\mathbf{c o r}}=\left(\frac{\partial X_{m}}{\partial t}\right)_{\mathbf{u n c o r}}+\left(\frac{\partial X_{m}}{\partial t}\right)_{\mathrm{n} d g}, 
\end{equation} 
where $(\partial X_{m}/\partial t)_{uncor}$ and  $(\partial X_{m}/\partial t)_{ndg}$ denote the before-nudged state and nudging tendency, respectively. The $(\partial X_{m}/\partial t)_{cor}$ represents the corrected stated of the state variable $X_{m}$, which should be close to ERA5. By construction, \eqref{eq3} tries to converge prediction from model towards true observations.

\subsection{Architecture of Deep Operator Network }
To learn a surrogate for prediction of the nudging tendency requires learning a function $\mathcal{G}$ from space $\mathbf{U}$ (before nudge data) to $\mathbf{V}$ (Nudging tendency),  a function of spatio-temporal coordinates $(x,y, t)$. DeepONet is a neural operator regression technique based on the universal approximation theorem \cite{chen1995universal}. The goal of DeepONet is to learn the mapping from an input space, $\mathbf{U(\eta)}$, to an output space $\mathbf{V(\zeta)}$.The conventional DeepONet architecture consists of two neural networks first, branch network (branch-net) and second trunk network (trunk-net).
The branch-network and trunk-network take in as input samples $u(\eta)$ $\in$ $\mathbf{U(\eta)}$ and $\zeta$, respectively. The output of the DeepONet can be represented as the operator $\mathcal{G}: \mathbf{U} \xrightarrow{} \mathbf{V}$, such that $v(\zeta) = \mathcal{G}(u(\zeta))$, where $u(\zeta)$ $\in$ $\mathbf{U}$ and $v(\zeta)$ $\in$ $\mathbf{V}$. The nudging tendency has many low energy modes, less coherent structures, and are very high dimensional due to which they are very hard to approximate using neural networks. This is due to very low correlations between feature and label data.  Oommen et. al \cite{oommen2022learning} has shown that these type of the problems can be learned successfully by using DeepONet  in latent space discovered by encoders. In this study, we use two autoencoders to reduce the spatial dimension of $\mathbf{U}$ and $\mathbf{V}$  to represent the nudging tendency in a smooth latent manifold so that its easier to learn. The steps required to train the  DeepONet are
\begin{enumerate}
    \item[i.] The first step is to train two autoencoders, one for the DeepONet input, i.e., the before-nudged variables ($\psi_{bf}(z,\bar{x},\bar{y},t)$) and the other for the DeepONet output that is the nudging tendency ($\psi_{ndg}(z,\bar{x},\bar{y},t)$ variables. Both autoencoders are trained using the MSE loss as shown:
    \begin{align*}
    \mathcal{L}_{AE1} &= ||\psi_{bf}(z,\bar{x},\bar{y},t) -\psi_{bf}(z,\bar{x},\bar{y},t;\theta_{AE1})||_{2}, \\
    \mathcal{L}_{AE2} &= ||\psi_{ndg}(z,\bar{x},\bar{y},t) -\psi_{ndg}(z,\bar{x},\bar{y},t;\theta_{AE2})||_{2},
\end{align*} 
where AE1 and AE2 represent the autoencoders for the before-nudged and the nudging tendency variables. Also, $\theta_{AE1}$ and $\theta_{AE2}$ represent the corresponding trainable parameters for the respective autencoders.
\begin{figure}[!ht]
    \centering
    \includegraphics[width=0.47\textwidth]{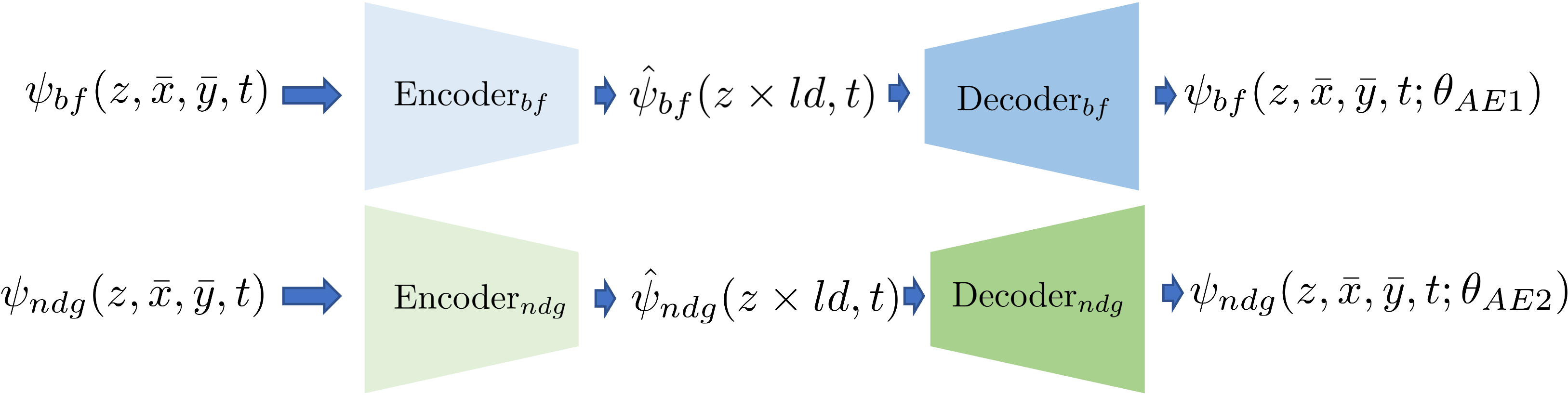}
    \caption{A block diagram showing high level view of the autoencoder. Here $\psi_{bf}(z,\bar{x},\bar{y},t)$ and $\psi_{ndg}(z,\bar{x},\bar{y},t)$ are the input to the encoders $Encoder_{bf}$ (encoder for before nudged data) and $Encoder_{ndg}$ (encoder for nudging tendency) respectively. $\hat{\psi}_{bf}(z \times ld,t)$, $\hat{\psi}_{ndg}(z \times ld,t)$ are the latent dimension representation of the respective inputs and $\psi_{bf}(z,\bar{x},\bar{y},t;\theta_{AE1})$, $\psi_{ndg}(z,\bar{x},\bar{y},t;\theta_{AE2})$ are the reconstructed variable via the $Decoder_{bf}$ (decoder for before nudged data) and $Decoder_{ndg}$ (decoder for nudging tendency) respectively. }
    \label{fig:E}
\end{figure}

\item[ii.] After training and validation of the encode and decoder, we  learn DeepONet in the latent space with features and labels defined by $\hat{\psi}_{bf}(z,\bar{x},\bar{y},t)$ and  $\hat{\psi}_{ndg}(z \times ld,t)$, respectively.  
The output of DeepONet $\hat{\psi}_{ndg\_don}(z \times ld,t)$ is trained against the  $\hat{\psi}_{ndg}(z \times ld,t)$ by minimizing the mean square error function expressed as
\begin{equation}
    \mathcal{L} = ||\hat{\psi}_{ndg\_don}(z \times ld,t) -\hat{\psi}_{ndg}(z \times ld,t)||_{2}.
\end{equation}
    
\begin{figure}[!ht]
    \centering   \includegraphics[width=0.45\textwidth]{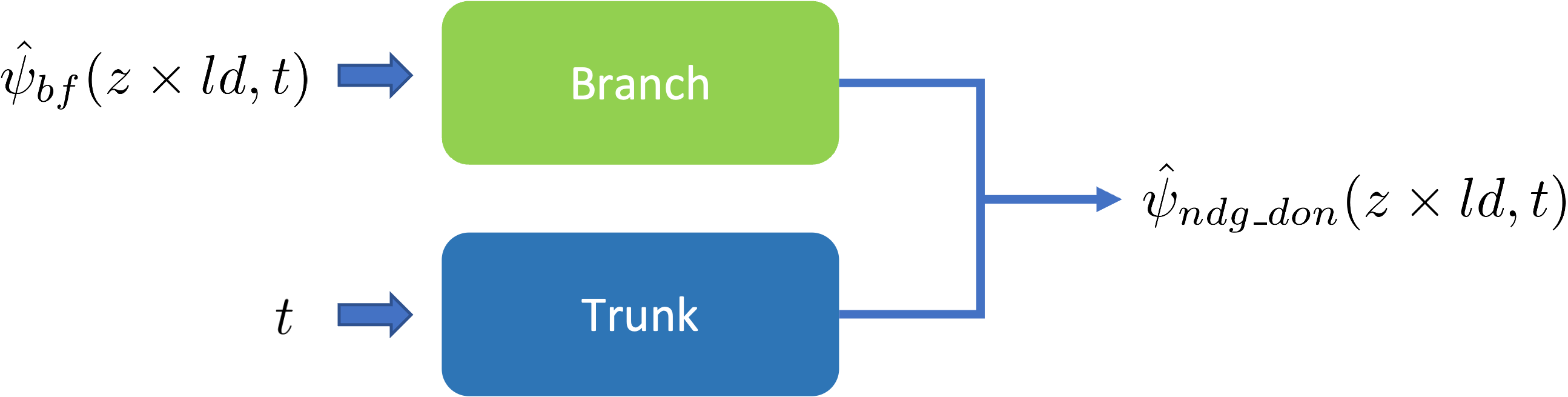}
    \caption{A block diagram showing a high level view of the DeepONet. $\hat{\psi}_{bf}(z \times ld,t)$ (latent space representation of before nudged state) is the input for the branch network and $t$ is the time which the input to the trunk network, which is the time for which we want to predict the nudging tendency $\hat{\psi}_{ndg\_don}(z \times ld,t)$.}
    \label{fig:D}
\end{figure}

\item[iii.] To recover the output of DeepONet in primitive space, we compose it with the decoder.
\begin{figure}[!ht]
    \centering
    \includegraphics[width=0.45\textwidth]{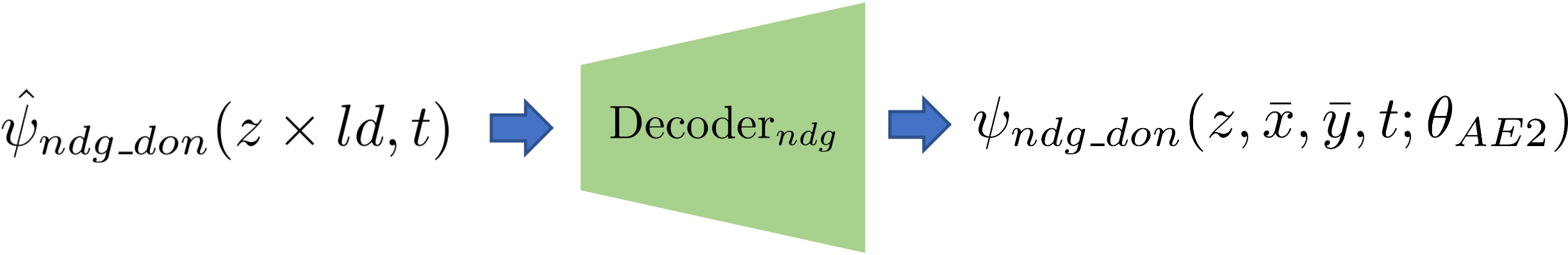}
    \caption{A block diagram showing a high level view of the decoding process after DeepONet prediction. Once the prediction of the nudging tendency is obtained in the latent space, it is passed through the $Decoder_{ndg}$ to get the primitive space representation $\psi_{ndg\_ don}(z,\bar{x},\bar{y},t;\theta_{AE2})$.}
    \label{fig:De}
\end{figure} 
\end{enumerate}

\begin{algorithm}[!ht]
\caption{AE based DeepONet algorithm}
\label{alg:algorithm}
\textbf{Input}: $\psi_{bf}$, $\psi_{ndg}$, $t$\\
\textbf{Parameter}: $\theta_{AE1}$, $\theta_{AE2}$, $\theta_{DeepONet}$ \\
\textbf{Output}: $\psi_{ndg\_don}$
\begin{algorithmic}[1] 
\STATE $\psi_{bf}\rightarrow AE1 \rightarrow \hat{\psi}_{bf}$
\STATE $\hat{\psi}_{bf}, t \rightarrow DeepONet \rightarrow  \hat{\psi}_{ndg\_don}$
\STATE $\psi_{ndg\_don}\rightarrow AE2 \rightarrow \psi_{ndg\_don}$
\STATE \textbf{return} $\psi_{ndg\_don}$
\end{algorithmic}
\end{algorithm}

\begin{table}[!ht]
\centering
\begin{tabular}{llllll}
 \hline
 \hline
 & Layer  & Feature  & Kernel  & Activation  \\
 \hline
 \hline
 & Conv2D  & 16  & $(3,3)$  & $ReLU$   \\
 & maxpool &  & $(2,2)$  &   \\
 & Conv2D & 8 & $(3,3)$  & $ReLU$  \\
 & maxpool &  & $(2,2)$  &   \\
 & Conv2D & 4 & $(3,3)$  & $ReLU$   \\
 & maxpool &  &$(2,2)$  &   \\
 & Flatten &  &  &   \\
 & Dense &  &  &   \\
 \hline
 \hline
\end{tabular}
\caption{Table representing architecture details of the encoder.}
\end{table} 

\begin{table}[!ht]
\centering
\begin{tabular}{llllll}
 \hline
 \hline
 & Layer  & Feature  & Kernel  & Activation  \\
 \hline
 \hline
 & Dense &  &  & $ReLU$  \\
 & Reshape &  &  &   \\
 & Transpose Conv2D & 4 & $(3,3)$  & $ReLU$   \\
 & Transpose Conv2D & 8 & $(3,3)$  & $ReLU$  \\
 & Transpose Conv2D  & 16  & $(3,3)$  & $ReLU$   \\
 & Conv2D & 4 & $(3,3)$ & \\
 & Reshape &  &  & \\
 & Dense &  &  &   \\
 \hline
 \hline
\end{tabular}
\caption{Table representing architecture details of the decoder.}
\end{table}
\begin{table}[!ht]
\centering
\begin{tabular}{llll}
\hline
\hline
& Branch-net &  \\
\hline 
\hline
 & Layer  & Activation  \\
 \hline
 \hline
 & Conv2D (feature=32,K=(3,3))  & $ReLu$  \\
  & Conv2D (feature=16,K=(3,3))  & $ReLu$  \\
 & Flatten   &  \\
 & Linear(ld)  & tanh \\
  & Linear(32)  & tanh \\
   & Linear(32) & tanh \\
     & Linear(32) & tanh \\
 & Linear($M \times ld$)   & \\
\hline
\hline
& Trunk-net &  \\
\hline 
\hline
& Linear(32) &  tanh \\
  & Linear(32) & tanh \\
   & Linear(32) & tanh \\
     & Linear(32)  & tanh \\
 & Linear($M \times ld$)  & \\
\hline
\hline
\end{tabular}
\caption{Table representing the DeepONet architecture.}
\end{table}

\section{Results}
To test our model, we train the DeepONet for a particular sub-region of the globe for the state variables $U$(zonal wind) and $V$ (meridional wind). The latitude and longitude of the chosen sub region varies  from 10N to 80N and 120W to 50W , respectively. The choice of the sub region is due to the larger variability of the atmospheric state $U$ and $V$ and is dominated by historical cyclones and storms. The DeepONet model is trained for the time span of October, which covers the famous hurricane Sandy in 2012. The DeepONet based model accurately predicts the nudging tendencies with a correlation higher than 0.7 with a confidence of 95\%. A correct prediction of the storm track of Sandy shows the robustness of the DeepONet model. 

In Figure 5, top/bottom left and top/bottom right sub-figures represent E3SM and DeepONet predicted nudging tendencies for $U$(zonal wind) and $V$(meridional wind), respectively.

\begin{figure}[!ht]
\centering
 \caption*{U - (Zonal wind)}
   \includegraphics[width=0.45\textwidth]{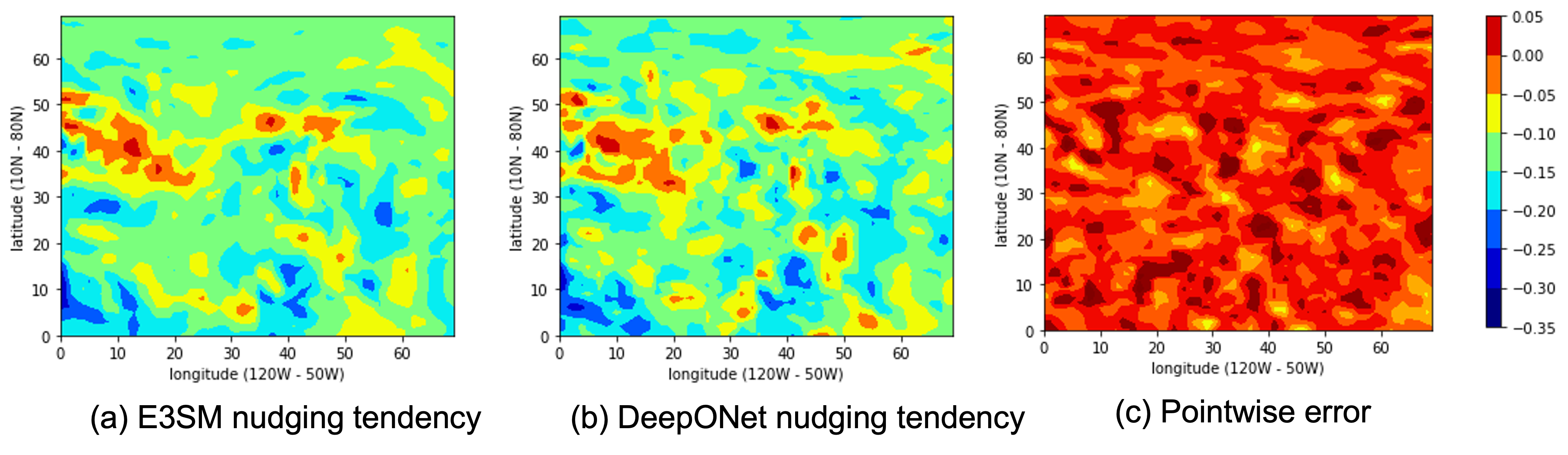}
   \label{fig:u1} 
    \caption*{V - (Meridional wind)}
   \includegraphics[width=0.45\textwidth]{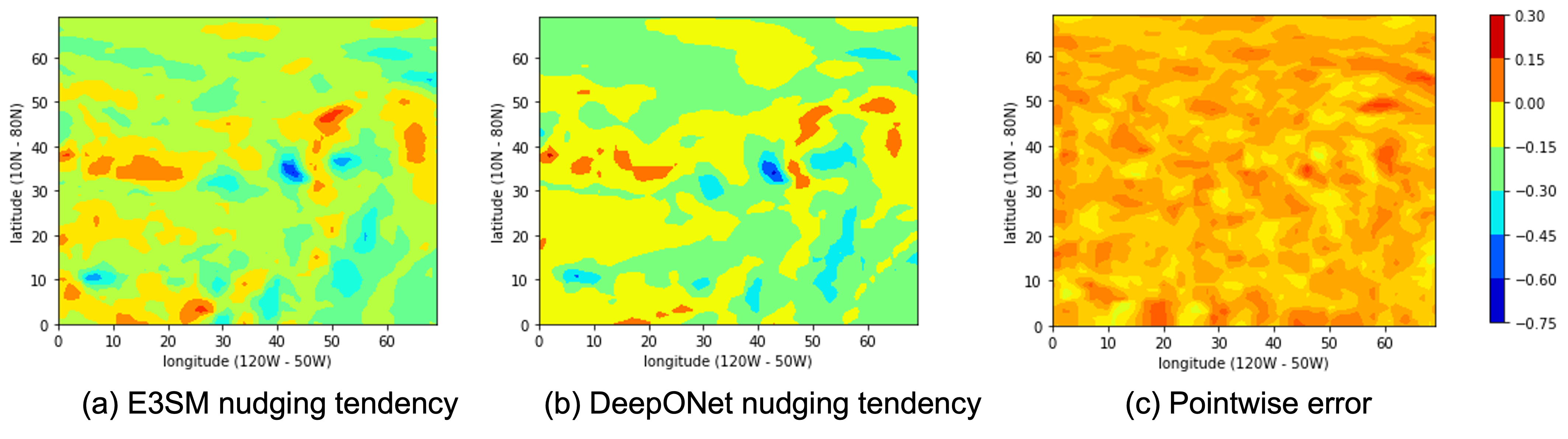}
   \label{fig:v1}
\caption{Figure showing nudging tendency predicted by DeepONet for U (zonal wind) and V (meridional wind) and compared against the corresponding E3SM nudging tendency in the test domain. Column (a) shows the E3SM nudging tendency, (b) shows the DeepONet predicted nudging tendency, and (d) shows the pointwise error for U and V respectively.}
\end{figure}
\begin{figure}[!ht]
\centering
    \caption*{U - (Zonal wind)}
   \includegraphics[width=0.45\textwidth]{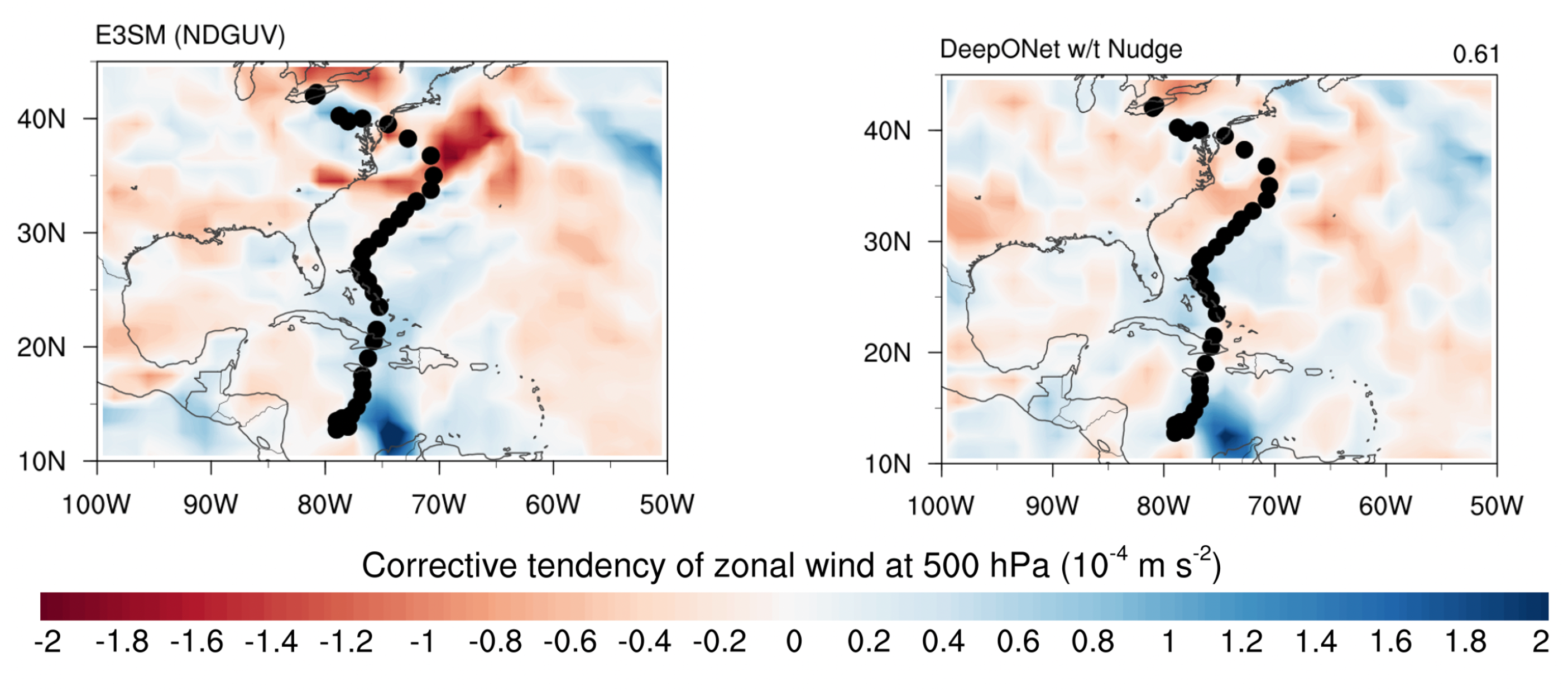}
   \label{fig:Q_train} 
   \caption*{V - (Meridional wind)}
   \includegraphics[width=0.45\textwidth]{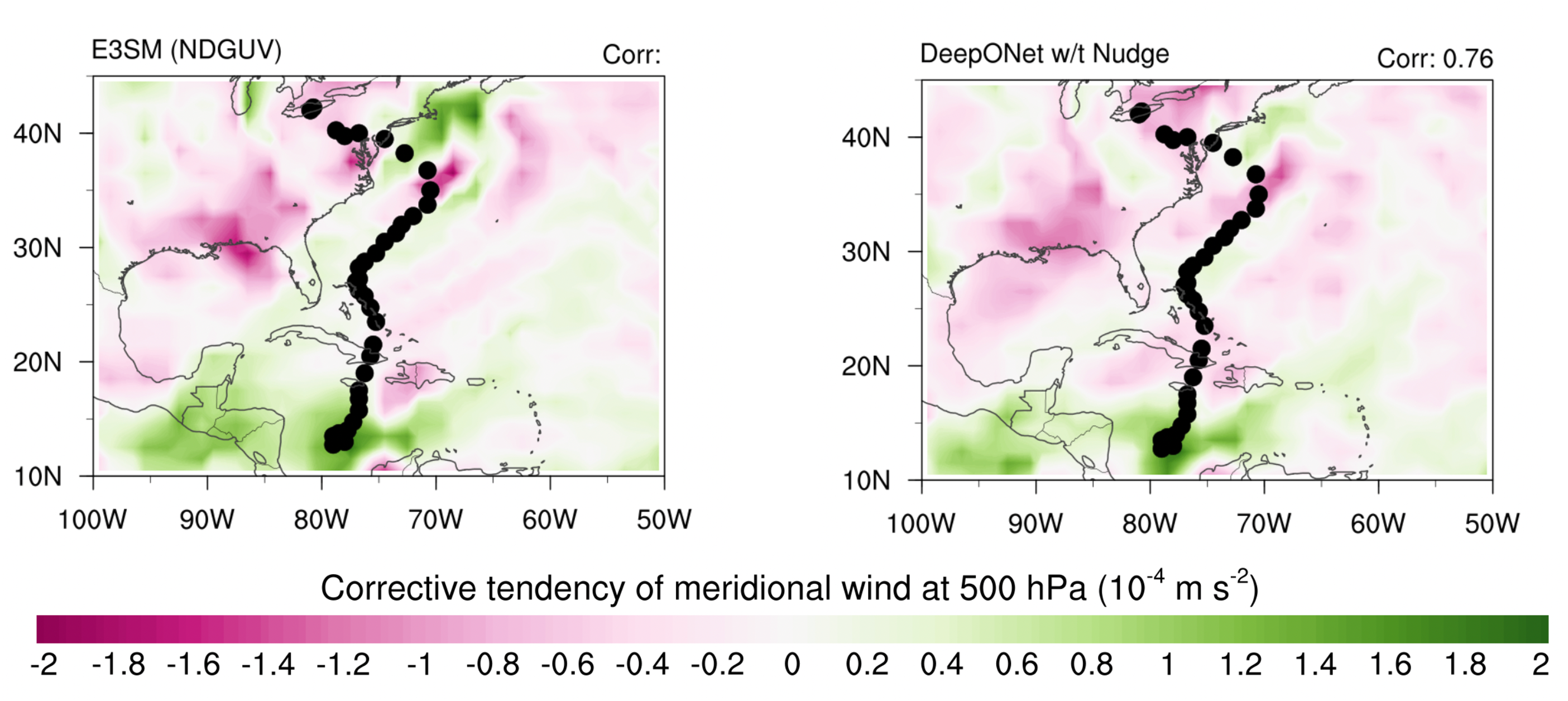}
   \label{fig:Q_test}
\caption{Figure showing the track of hurricane sandy in 2012. The left column sub-figures are from the E3SM nudged UV simulations and the right column sub-figures are the DeepONet simulation trained with nudging tendency for $U$(zonal wind) and $V$(meridional wind) respectively.}
\end{figure}
\section{Summary}
In this work, we developed a DeepONet based surrogate model, which successfully learns the bias correction in climate models. We first trained two auto-encoders for $U$ and $V$ providing their latent representations. The evolution of latent representation of $U$ and $V$ are learned by DeepONet.
 Latent representation of features and labels helps DeepONet to learn the system dynamics fast and efficiently.
 The accuracy of DeepONet modes is established by testing it to predict the path of hurricane Sandy. 
 The ongoing efforts include integration of the trained DeepONet model in the E3SM, so that it can be used to correct the biases in an online mode for more skillful simulations. 

\section{Acknowledgments}
The authors would like to thank DARPA for funding this work through the project AI-assisted Climate Tipping-point Modeling (ACTM), Agreement No. HR00112290029. K.S. and A.B would also like to acknowledge Vivek Oommen (Brown University). K.S. and A.B would also like to acknowledge the use of the CCV's Oscar HPC cluster at Brown University.

\bibliography{aaai22}
\end{document}